\documentclass[%
 reprint,
superscriptaddress,
preprintnumbers,
 amsmath,amssymb,
 aps,
prb,
floatfix,
]{revtex4-1}

\usepackage{graphicx}
\usepackage{dcolumn}
\usepackage{bm}

\usepackage{times}

\begin{document}

\title{Designing Disorder into Crystalline Materials}

\author{Arkadiy Simonov}
\affiliation{Department of Materials, ETH Zurich, CH-8093 Zurich, Switzerland}

\author{Andrew L. Goodwin}
\affiliation{Department of Chemistry, University of Oxford, South Parks Road, Oxford OX1 3QR, U.K.}

\date{\today}

\begin{abstract}
Crystals are a state of matter characterised by periodic order. Yet crystalline materials can harbour disorder in many guises, such as non-repeating variations in composition, atom displacements, bonding arrangements, molecular orientations, conformations, charge states, orbital occupancies, or magnetic structure. Disorder can sometimes be random, but more usually it is correlated. Frontier research into disordered crystals now seeks to control and exploit the unusual patterns that persist within these correlated disordered states in order to access functional responses inaccessible to conventional crystals. In this review we survey the core design principles at the disposal of materials chemists that allow targeted control over correlated disorder. We show how these principles---often informed by long-studied statistical mechanical models---can be applied across an unexpectedly broad range of materials, including organics, supramolecular assemblies, oxide ceramics, and metal--organic frameworks. We conclude with a forward-looking discussion of the exciting link to function in responsive media, thermoelectrics, topological phases, and information storage.
\end{abstract}

\maketitle


\section{Introduction}

All materials are disordered at finite temperature. In crystals, where this disorder is usually dominated by thermal motion, the equilibrium atomic positions are themselves periodic and hence ordered. Departures from this paradigm are well known and have been studied since the very earliest days of structural science.\cite{Warren_1934,Pauling_1935,Cartwright_2012} Liquids, glasses, and gels possess no long-range structural periodicity; fibres and liquid crystals are ordered in fewer than three dimensions; microporous materials such as zeolites and metal--organic frameworks (MOFs) may themselves be crystalline but can nonetheless contain a fluid- or glass-like guest phase within their pores. But even \emph{bona fide} crystals can be disordered in ways other than by thermal motion. This happens whenever there is some component with an internal degree of freedom. Paramagnetic MnO is still a crystal even though the Mn$^{2+}$ magnetic moments are not ordered; in plastic crystals---as in ices---molecular centres-of-mass exhibit long-range order but molecular orientations do not; alloys are positionally ordered yet compositionally disordered [Figure 1a--c]. Such are the disordered crystals of seminal texts in the field,\cite{Parsonage_1978,Ziman_1979} and the focus of this review.

\begin{figure}
\begin{center}
\includegraphics{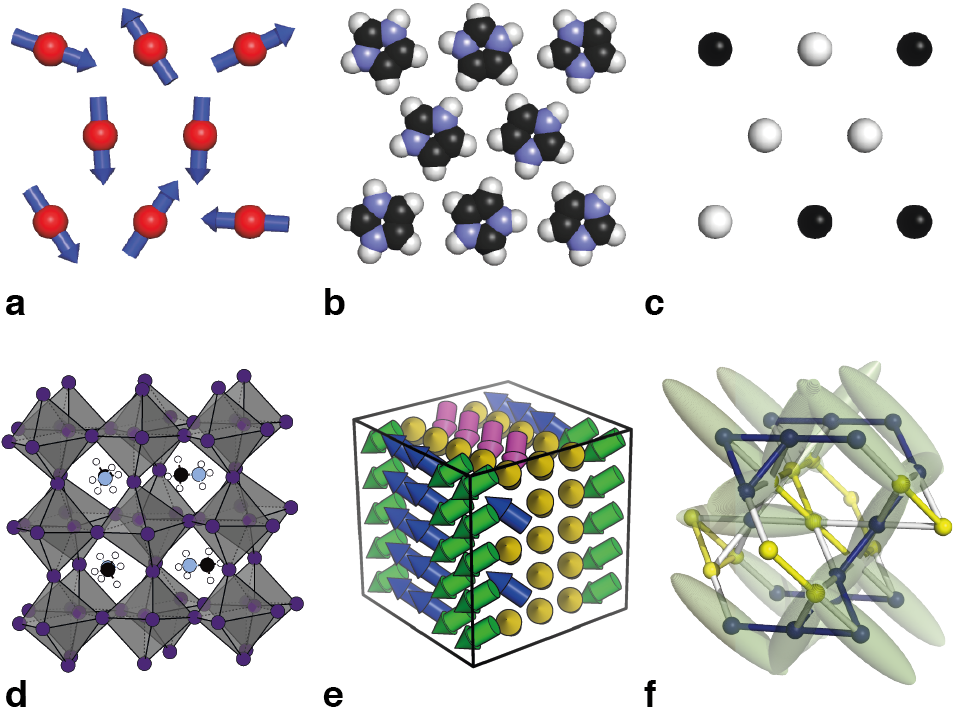}
\end{center}
\caption{\label{fig1} {\bf Canonical disordered crystals.} (a) In a paramagnet, the positions of magnetic ions (red spheres) is periodic, but the arrangements of the corresponding spins (blue vectors) are not. (b) Plastic crystals are similar, with the disorder now involving molecular orientations. (c) Disordered alloys contain a superposition of positional order and compositional disorder. (d) The photovoltaic lead--halide perovskites support a variety of disorder types, including orientational disorder of methylammonium cations (C = black, N = blue, H = white spheres) and distortions of the anionic lead--halide framework (connected polyhedra).\cite{Zhu_2019} (e) In ferroelectric BaTiO$_3$, polarisation emerges from correlated disorder of Ti$^{4+}$ off-centering along local $\langle111\rangle$ axes (coloured arrows).\cite{Senn_2016} (f) And the emergence of ferromagnetism in Fe$_3$O$_4$ couples to local Fe$^{2+}$/Fe$^{3+}$ charge order in a dynamically-disordered `trimeron liquid' phase (Fe$^{2+}$ in yellow, Fe$^{3+}$ ion blue, and trimerons as shaded ellipsoids).\cite{Perversi_2019}}
\end{figure}


Our motivation for surveying the ways in which disorder might be incorporated intentionally within crystals and its effect on their physical properties comes from the empirical finding that many functional crystalline materials are indeed disordered in some non-trivial way. The photovoltaic hybrid lead--halide perovskites,\cite{Weller_2015} ferroelectric BaTiO$_3$,\cite{Comes_1970,Senn_2016} and magnetite\cite{Perversi_2019} (Fe$_3$O$_4$) are just some examples of recent or enduring interest [Figure 1d--f]. In each case it is clear that the disordered component---whatever its particular nature---is far from random, and the presence of specific correlations or patterns is likely important for the particular function of interest. Of course the nature and phase behaviour of disordered crystalline states have long been studied within the statistical mechanical and crystallographic communities.\cite{Onsager_1944,Ziman_1979,Welberry_1985} So too are the physical properties of functional materials the bread-and-butter of condensed matter physics and materials chemistry. Yet the \emph{link} between correlated disorder and property is an emerging, active, and challenging field at the very frontier of structural science. At its heart lie a number of important questions. How does one design a disordered crystal? What is the form of that disorder, how is it controlled, and how is it characterised? And---most importantly---what properties does it impart? In the answers to these questions lie the secret of how one might design functional materials by controlling disorder.

Because correlated disorder is often a consequence of geometry, its nature can transcend the particular chemistry or physics from which it evolves. Hence there exist unexpected parallels between ostensibly unrelated materials: magnetocaloric Tb(HCOO)$_3$ and benzenetrisamide nucleating agents, or PbMg$_{1/3}$Nb$_{2/3}$O$_3$ relaxors and Prussian Blue analogue cathode materials. While each system has its own peculiarities---such as we will come to discuss in detail below---there are nonetheless common aspects that influence them all. For example, each system has access to a large configurational degeneracy that in turn allows a heightened response to external stimuli. Likewise, the particular patterns that persist within a disordered state are characterised by specific periodicities; these can couple to other materials properties that also depend on periodicity, such as electronic states and collective vibrations (\emph{i.e.}\ phonons).\cite{Overy_2016} While our focus is on materials where disorder arises from atomic-scale degrees of freedom, the underlying geometric origin of this disorder means one can often develop mappings that span lengthscales. Hence a design strategy for manipulating electronic band structure, say, may be as relevant to photonics when translated onto optical length scales.\cite{FroufePerez_2016} And, reversing the argument, understanding how shape drives disordered packings of (macroscopic) polyhedra can in turn help direct molecular orientational disorder in hybrid materials.\cite{Damasceno_2012}


The bulk of our review is arranged according to a series of core chemical design strategies for incorporating particular types of disorder. These involve electronic instabilities, compositional complexity, directed self-assembly, molecular shape, and low-energy dynamics. In each section we find ourselves drawing on a handful of paradigmatic statistical mechanical models that have largely been developed in the context of frustrated magnetism. Consequently we pre-empt our survey of design strategies by covering the basics of these models, placing them as much as possible in a chemical context. As we then take each design strategy in turn, we illustrate its relevance to a number of examples drawn from the recent literature. Our hope is to make clear the role of geometry and the link to function wherever possible. The review concludes by looking forward: we attempt to anticipate areas of materials chemistry in which disordered crystals may offer uniquely valuable opportunities for functional materials design and discovery.

\section{Models of correlated disorder}
{\bf Degrees of freedom.} The three fundamental ingredients for statistical mechanical models of disordered crystals are (i) the relevant microscopic degrees of freedom, (ii) the interactions between those degrees of freedom, and (iii) the lattice on which these are collectively arranged.\cite{Ziman_1979} Much of the corresponding literature is phrased in terms of spin models, but the parallels to chemical systems are often straightforward and usually well established.\cite{Parsonage_1978} So, for example, Ising degrees of freedom would, in a magnetic context, describe magnetic spins that can orient parallel or antiparallel to some axis.\cite{Ising_1925} But the corresponding models are entirely unaffected by the physical origin of the Ising degree of freedom, and hence are universal to any system where components can be in one of two distinguishable but equivalent states.\cite{Brush_1967} Some chemical examples are given in Figure 2a, which also includes examples for various more complex degrees of freedom. A natural extension of the Ising model, for example, is to allow any one of $n>2$ states; this gives the $n$-state Potts models.\cite{Potts_1952} The three possible orientations of a dimethylammonium cation (DMA$^+$) on the 12-coordinate site of the hybrid perovskite [DMA]Mn(HCOO)$_3$ or the various orientations of Mn$^{3+}$ Jahn-Teller distortions in LaMnO$_3$ are two examples that have been interpreted in terms of Potts models.\cite{Simenas_2016,Ahmed_2006} Mathematically, all Ising and Potts degrees of freedom behave as discrete scalars.

\begin{figure*}[t]
\begin{center}
\includegraphics{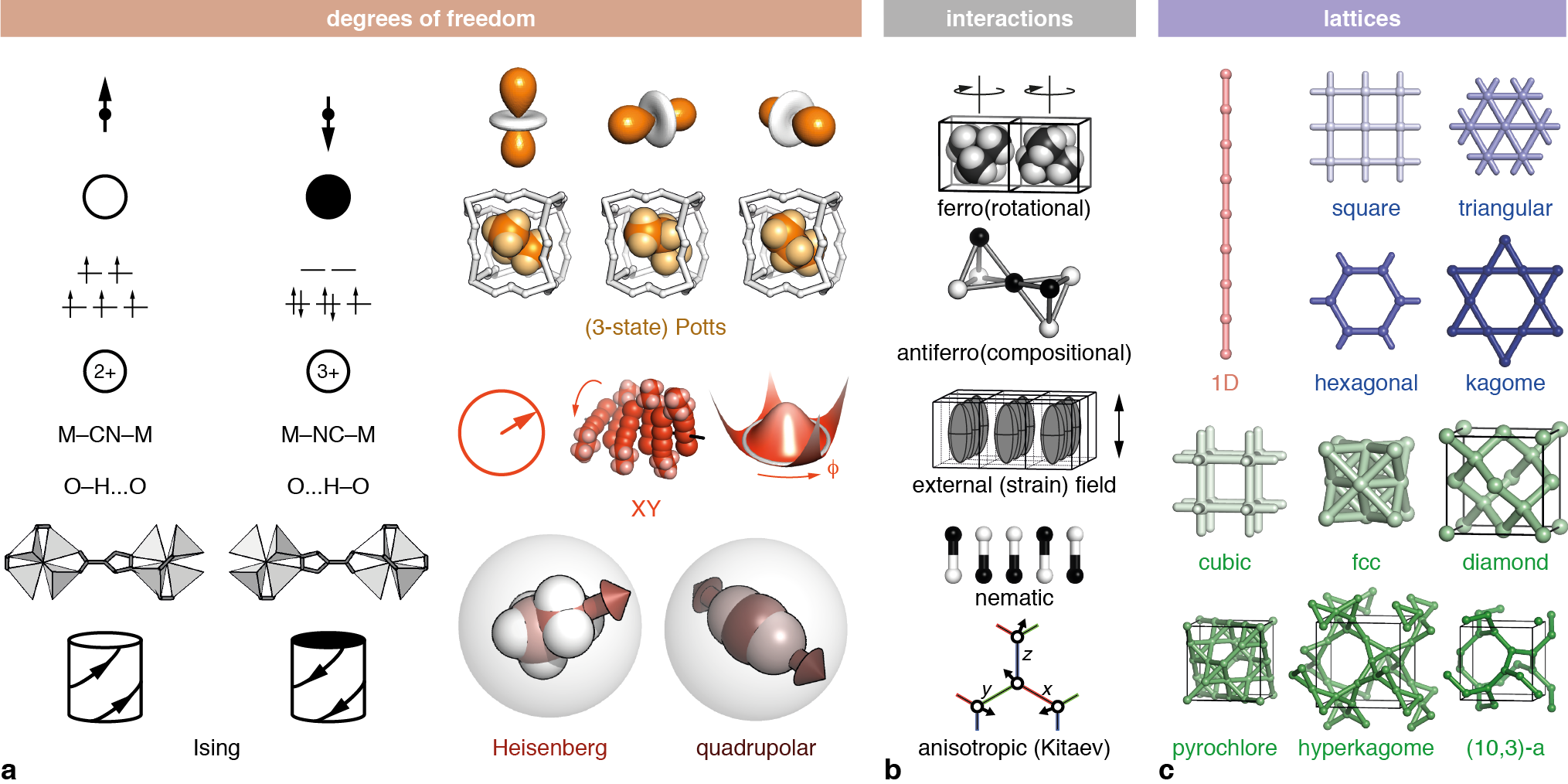}
\end{center}
\caption{\label{fig2} {\bf Common local degrees of freedom, interactions, and lattice types.} (a) Degrees of freedom may be discrete, such as the binary Ising (black) and $n$-state Potts (orange) models. Commonly-encountered continuous degrees of freedom include XY states (red) and Heisenberg or quadrupolar spins (brown). (b) Representations of some key interactions of relevance to disordered states. (c) High-symmetry lattices in one (pink), two (blue), and three (green) dimensions. All these lattices are both uninodal and edge-transitive. In other words, each site in a given lattice is related to all other sites by crystal symmetry; so too for the lattice edges (interaction pathways).}
\end{figure*}

The simplest \emph{continuous} degree of freedom is that of the XY models, where (in the magnetic case) spin vectors of unit length are confined to lie within a plane.\cite{Vaks_1966} Hence their orientation is described by a phase angle $0\leq\theta<2\pi$. A straightforward chemical analogue is the orientation of molecules in rotator phases,\cite{Sirota_1996} but there are also more subtle mappings to collective tilt distortions,\cite{Meier_2017} superconducting states,\cite{Abrikosov_1957} and vertical shifts in periodic columnar phases.\cite{Cairns_2016} If not all orientations $\theta$ are equally likely---perhaps as a result of the local crystal field---then the system is said to be anisotropic and the crystal free energy may include so-called `single-ion' terms taken from the series expansion $D_1\cos\theta+D_2\cos2\theta+\ldots$. In three dimensions (3D), the simplest degree of freedom is the Heisenberg spin: a unit vector that can point anywhere in space.\cite{Joyce_1967} The orientation of polar molecules, such as the MA$^+$ cations in MAPbI$_3$,\cite{Leguy_2015} or cation displacements, such as the off-centering of Ti atoms in BaTiO$_3$, can be described in this way (often with some significant anisotropy, now in 3D and hence characterised by the spherical harmonics).\cite{Zhong_1995} Heisenberg spins are to dipolar degrees of freedom (\emph{e.g.}\ displacement or orientation) as Ising states are to monopoles (\emph{e.g.}\ site occupancy or charge state). 

In turn, one might anticipate the relevance of quadrupolar and higher-order multipolar equivalents. These have indeed be used to describe the phase behaviour of disordered crystals, where the key multipole is usually determined by the shape of a relevant molecule; for example, octupoles for tetrahedral molecules such as CH$_4$ or hexapoles for trigonal-planar molecules such as the nitrate anion.\cite{James_1959} Formally, these more complex degrees of freedom are characterised by tensors, the dimensions of which depend on the order of the corresponding multipole. Quadrupoles are usually the most straightforward to interpret, as they can be represented by $3\times3$ matrices and relate to the orientation of nonpolar rod- or disc-like molecules, such as CO$_2$ or the guanidinium cation. The corresponding statistical mechanics is that of nematic phases, which include an important class of liquid crystals.\cite{Frenkel_1991} There are other, less frequently encountered, degrees of freedom of relevance to disordered crystals. Examples include the use of quaternions to describe the orientation of molecules of arbitrarily low symmetry and axial vectors for freely-rotating objects.\cite{Karney_2007} 

{\bf Interactions.} Long-range interactions tend to drive long-range order. It is unsurprising therefore that the key interactions between degrees of freedom in disordered crystals are usually short-range in nature [Figure 2b]. The simplest and most common involves the inner (scalar) product taken across neighbouring sites: this is a general measure of the similarity of the two degrees of freedom at these sites, irrespective of the particular mathematics involved. For example, in the case of two Ising states $e_i,e_j=\pm1$, the inner product $e_ie_j$ is (obviously) $+1$ if $e_i=e_j$ and $-1$ otherwise; for XY or Heisenberg degrees of freedom $\mathbf S_i,\mathbf S_j$ the inner product $\mathbf S_i\cdot\mathbf S_j$ is again obviously largest if $\mathbf S_i=\mathbf S_j$ and has the same functional form as the exchange interaction for magnetic spins. For this reason, the interaction is often called exchange (or `effective' exchange) irrespective of its physical origin; its strength is usually given the symbol $J$.\cite{Ziman_1979} The sign of $J$ determines whether neighbouring sites tend to adopt similar or dissimilar states---these cases being the equivalents of ferromagnetic or antiferromagnetic exchange. The prefixes `ferro' and `antiferro' are borrowed widely to indicate the sign of the interaction. So the MA$^+$ cations in [MA]Mn(HCOO)$_3$ interact in an antiferrodipolar sense, since their orientation alternates between neighbouring sites;\cite{PatoDoldan_2016} by contrast the guanidinium (Gua$^+$) cations in [Gua]Cd(HCOO)$_3$ exhibit ferroquadrupolar order since their plane normals align along a single common axis.\cite{Evans_2016}

There are a number of possible variations on this theme. More distant interactions may be important, such as those between next-nearest neighbours; in such cases one ordinarily expects a reduction in the magnitude of the corresponding $J$. This is chemically sensible, since the steric interactions between distant molecules are indirect and likely rely only on mediation by the surrounding lattice. Some long-range interactions can adopt the same effective form as an exchange interaction. The best known example is that of dipoles on the pyrochlore lattice: a geometric peculiarity is that the contribution of next-nearest-neighbour (and beyond) interactions effectively cancel, and the surviving nearest-neighbour term can be recast in terms of the inner product $\mathbf S_i\cdot\mathbf S_j$.\cite{denHertog_2000} Interactions with an external field usually take the same form as exchange, except that the coupling constant $J$ is in the general case a tensor, the dimensions of which depend on the natures of the relevant degree of freedom and conjugate field.


Another variation involves higher-order exchange, such as the biquadratic interaction for XY spins, which is proportional to $(\mathbf S_i\cdot\mathbf S_j)^2$. In spin models these terms control the degree of nematic order---the tendency for spins to align along a common axis, whether parallel or antiparallel.\cite{Lauchli_2006} A chemical example is that of solid CO, where the alignment axes of molecules shows long-range order, but there is head-to-tail disorder of the CO dipoles.\cite{Lipscomb_1974} We will come to discuss a less obvious mapping in the mixed-metal cyanides Au$_x$Ag$_{1-x}$(CN) in more detail below. The simplest picture is that these interactions represent the second-order term in the general series expansion $J(\mathbf S_1,\mathbf S_2)=J_1(\mathbf S_1\cdot\mathbf S_2)+J_2(\mathbf S_i\cdot\mathbf S_j)^2+\ldots$. Yet further variations are that of antisymmetric exchange,\cite{Dzyaloshinsky_1958,Moriya_1960} and of multi-body interactions---\emph{i.e.}\ involving more than two sites---such as the six-body terms thought to govern cation order in some rocksalt sufides.\cite{UronesGarrote_2005}

A final interaction type we will need to consider is that of the compass models.\cite{Nussinov_2015} These involve anisotropic interactions, by which we mean that neighbouring sites interact differently depending on their mutual orientation. Probably the most famous example is the Kitaev interaction on the honeycomb lattice.\cite{Kitaev_2006} Honeycombs involve three different edge orientations offset by 120$^\circ$; in the Kitaev model, Heisenberg spins connected by one type of edge interact \emph{via} their $x$-components, those connected by a second type \emph{via} their $y$ components, and the third the $z$. Kitaev was primarily concerned with the properties of the quantum ($S=\frac{1}{2}$) case, and indeed his approach has been used not only to study magnetic disorder but also to describe collective orbital states in strongly-correlated oxides.\cite{Nussinov_2015} We will come to show that simple classical analogues emerge naturally (if surprisingly) from sensible bonding considerations in ferroelectrics and metal--organic frameworks. The basic idea will be much the same: the $x$-components of the local degrees of freedom affect only those neighbours connected along one subset of axes, and so on.

{\bf Lattice.} It is self-evident that lattice geometry will be crucial to the phase behaviour of Kitaev-like systems, since interaction and orientation are so explicitly linked. But even for simpler interaction types one intuitively (and rightly) expects different phase behaviour for different lattice topologies: the number of neighbours for each site may differ, and the connectivity will determine the extent to which neighbours of a given site are also neighbours of each other. Symmetry is also crucial. Since disorder is favoured by the existence of a shallow configurational landscape containing many equivalent or nearly-equivalent states, one tends to find correlated disordered states favoured by high-symmetry lattices. A handful of the most common examples is given in Figure 2c.


\noindent{\bf Geometric frustration and competing interactions.} Perhaps the key advantage of couching disordered phases in terms of their underlying statistical mechanics is that the phase behaviour of these various models is almost always known from theory (or, if not, from computation). It is a textbook result, for example, that 1D models do not order at any finite temperature, and hence any system whose behaviour maps onto 1D physics will itself fail to order. Likewise, it is known that Ising models can order in 2D, but those with continuous degrees of freedom are always unstable with respect to vortex formation.\cite{Kosterlitz_1973} Some models, such as the Heisenberg ferromagnets, are uncomplicated; others give particularly unusual states that are disordered but far from random.

A frequently-encountered example of the latter is the triangular Ising antiferromagnet (TIA). Its ground state is not unique but instead contains all those configurations for which individual triangles contain two `up' and one `down' Ising state, or \emph{vice versa} [Figure 3a].\cite{Wannier_1950} The system is said to be geometrically frustrated because of the incompatibility between its degrees of freedom and the lattice on which they sit (\emph{cf}\ the square Ising antiferromagnet, which is not frustrated). In physical realisations of the TIA, longer-range interactions---however weak---can act to break crystal symmetry and remove this frustration at some sufficiently low temperature $T_{\rm c}$. For the temperature regime $T_{\rm c}\leq T\lesssim |J|$ the system is disordered but strongly correlated; the ratio $f=|J|/T_{\rm c}$ is a crude measure of the degree of frustration ($f>10$ is typically considered strong). Geometric frustration plays a key role in stabilising the states of many disordered functional materials.\cite{Moessner_2006}

\begin{figure}
\begin{center}
\includegraphics{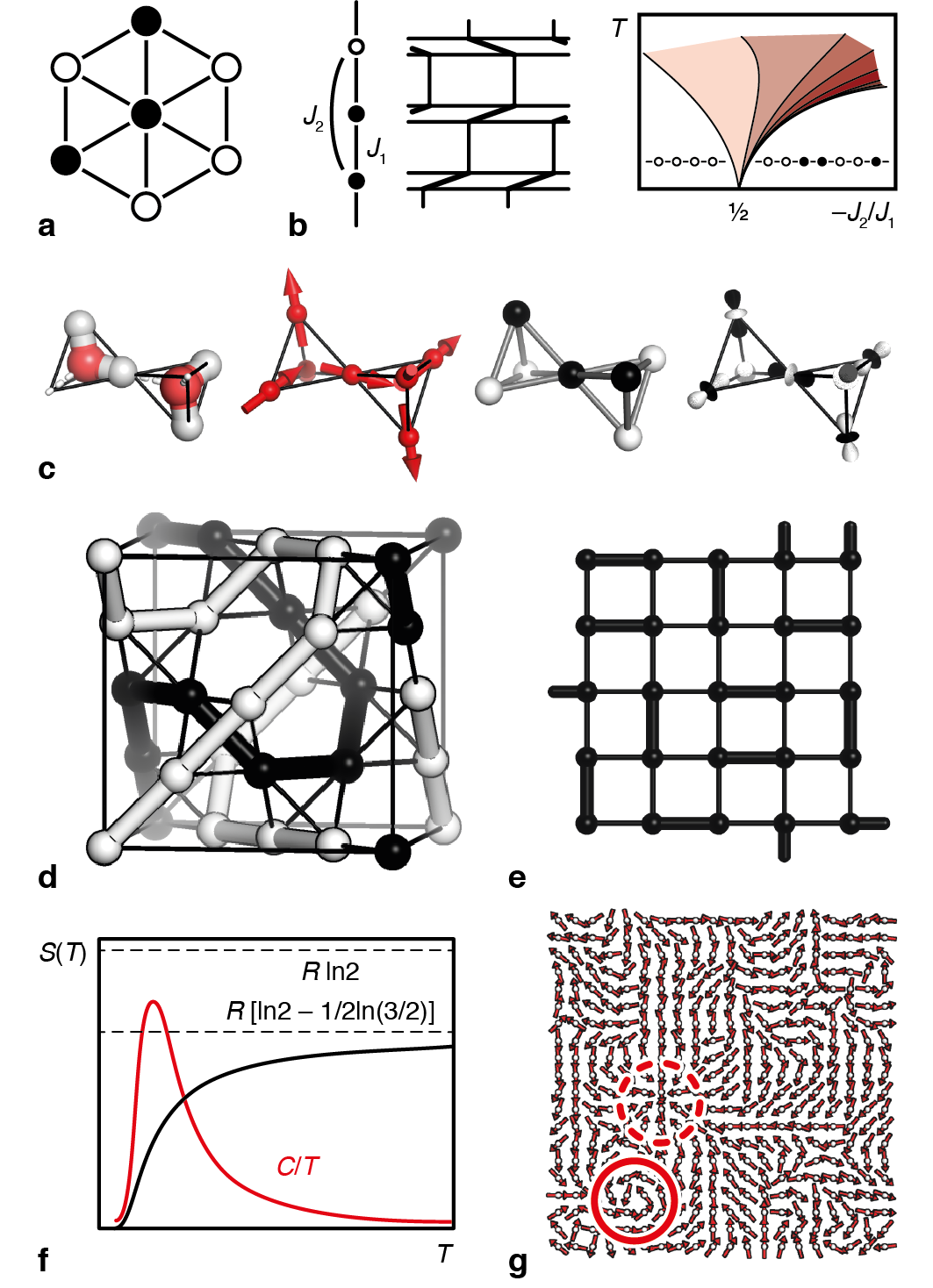}
\end{center}
\caption{\label{fig3} {\bf Models of disorder and competing interactions.} (a) The ground state of the triangular Ising antiferromagnet: note no triangle has at its vertices three Ising states of the same type. (b) Competing nearest-neighbour ($J_1$) and next-nearest neighbour ($J_2$) interactions in the 1D ANNNI model give a series of complex phases (shaded region of phase diagram) of relevance to stacking sequences in layered materials.\cite{Fisher_1980,Cheng_1988} (c) Ice-like states map to the ground states of the pyrochlore Ising antiferromagnet. Here, the vertices of each tetrahedron partition into two Ising states of one type, and two of the other. (d) Spaghetti phases are charaterised by dense Ising lattice decorations in which like Ising states form connected one-dimensional paths; as such, each site is associated  with two of its neighbours. (e) Dimer states occur when each site is associated with a single neighbour. (f) Transitions to correlated disordered states are usually characterised by a broad maximum in the specific heat (red line); the corresponding entropy change is characteristic of a given disordered state, such as that shown here (black line) for the spin-ice Dy$_2$Ti$_2$O$_7$.\cite{Ramirez_1999} 2D continuous models are susceptible to the formation of topological defects, such as vortex--antivortex pairs (filled circle; dashed circle).}
\end{figure}

Systems containing competing interactions can exhibit a similar kind of structural complexity to that seen in geometrically frustrated phases. Probably the best known example of this type is the 1D axial next-nearest neighbour Ising (ANNNI) model [Figure 3b].\cite{Bak_1982} There are two interactions in this model: $J_1$ operates between neighbouring sites, and $J_2$ between next-nearest neighbours. For certain choices of $J_1,J_2$ the two interactions are irreconcilable. Consider, for example, the case where $J_1$ favours like Ising states at neighbouring sites, but $J_2$ favours state inversion. Any triplet of three successive sites must violate at least one or other interaction. The phase behaviour that arises for different $J_1/J_2$ ratios is (in)famously complex; and, as we will explore in more detail below, ANNNI models have been used to rationalise the structural complexity of various families of layered materials.

\noindent{\bf Correlated disorder, hidden order, and emergence.} There are in principle a very large number of different disordered states accessible using the various degrees of freedom, interactions, and lattice types enumerated above. In practice, however, certain states recur more frequently than others. We have already flagged the TIA and ANNNI models as examples; Figure 3c--e highlights a handful of others. A particularly well known example in the chemical literature is the family of ices, all members of which are related to the pyrochlore Ising antiferromagnet.\cite{Pauling_1935,Bramwell_1998} As for the TIA, the ground state of this model is not unique but is characterised by a famous local rule---namely, that each tetrahedron of the pyrochlore lattice connects two sites in one Ising state and two in the other. In water ice itself the Ising states represent the directions of hydrogen bonds between neighbouring H$_2$O molecules. As this local rule propagates across the pyrochlore lattice, the pairs of neighbouring sites in a common Ising state form a dense set of non-intersecting 1D paths; this is a specific example of a broader family of so-called loop or spaghetti phases.\cite{Jaubert_2011,Baise_2018} Each site is a member of exactly one spaghetto because it shares its Ising state with exactly two neighbours---one in each of the two tetrahedral units of which it is a member. Dimer states are a related family, for which each site is associated with exactly one (rather than two) of their neighbours.\cite{Anderson_1973} These various unconventional states are different examples of `procrystals' that can be identified on the basis of their highly-structured diffuse scattering patterns.\cite{Overy_2016,Keen_2015}

From a conventional crystallographic viewpoint, there is no distinction between the symmetries of correlated disordered states and that of their random counterparts (\emph{i.e.}\ uncorrelated degrees of freedom). Indeed the transition from one to the other is often accompanied by a broad specific heat anomaly rather than the divergence associated with conventional phase changes; in some cases the transition is said to involve `hidden order'.\cite{Parsonage_1978} The entropy change associated with formation of the correlated state can of course be measured, and historically there was significant interest placed in relating its value to the microscopic degrees of freedom at play [Figure 3f]. Pauling's early rationalisation of the residual entropy of ice is a particularly successful example.\cite{Pauling_1935} Nevertheless we now know that, at least in some cases (most famously the ices), the correlated state is characterised by a very different type of symmetry---that of an emergent gauge field---to which diffraction measurements are sensitive via the diffuse scattering contribution.\cite{Henley_2010}






A final point we need to make is that many of the disordered states we consider are known to support emergent phenomena operating on a length-scale significantly removed from that of the fundamental degrees of freedom from which they arise. We have already alluded to existence of vortex states in some 2D models; these vortices and their 3D equivalents (\emph{e.g.} skyrmions and strings) are topological emergent objects that can carry a quantised effective charge, and in turn can be used to store information [Figure 3g].\cite{Muhlbauer_2009} Likewise, excitations of the gauge field symmetry found in ices behave as emergent monopoles that interact with one another according to an emergent electrostatics.\cite{Henley_2010} More generally, the excitations that transform any one specific disordered configuration into another identify emergent multi-body degrees of freedom that are intermediate to the localised excitations of isolated molecules and the collective phonon modes of conventional crystals.\cite{Oakes_2016} The key conclusion is that correlated disordered states are known \emph{a priori} to support a variety of unusual physical properties, many of which can be anticipated by understanding the statistical mechanics of the corresponding microscopic model. Inverting this link then into a design strategy: we can in principle hope to engineer materials with a specific emergent property by controlling the underlying degrees of freedom, interactions, and crystal lattice.


\section{Design strategies}

\noindent{\bf Electronic instabilities.} The Jahn Teller (JT) effect and its variants provide an immediately attractive design strategy for incorporating disorder in crystals and for linking that disorder to electronic properties.\cite{Pearson_1975,Goodenough_1998} JT distortions lower local symmetry, so there is automatically a family of equivalent JT states; interconverting amongst these then corresponds to a local degree of freedom. Since JT stabilisation involves redistribution of electrons and variation in bond lengths, there is a natural source of interactions between these degrees of freedom. And, because orbital occupancies are affected, different cooperative JT states often give different electronic and/or magnetic behaviour.

Taking the manganite perovskite LaMnO$_3$ as a canonical example, its $t^3_{2g}e^1_g$ Mn$^{3+}$ $d$-electron configuration is degenerate and so can be stabilised by coupling to a local distortion of the same $E_g$ symmetry.\cite{Goodenough_1998,Goodenough_1955} In principle, the twofold degeneracy of the $E_g$ configuration means that the available distortion space is itself two-dimensional. But in practice the subset of six discrete distortions giving a pair each of long, short, and intermediate Mn--O bonds (the so-called $Q_2$ state) is favoured energetically [Figure 4a]. This distortion couples O atom positions to the occupancy of $e_g$ orbitals. And, because each O atom connects two Mn$^{3+}$ centres, there is a natural mechanism for generating interactions between neighbouring distortions (and hence $e_g$ orbital occupancies) of the Mn$^{3+}$ sublattice. So the choice of local distortion orientation behaves as a Potts-type degree of freedom, and the cooperative JT effect provides a direction-dependent antiferroic coupling between neighbouring states (\emph{i.e.} long Mn--O bonds tend to avoid meeting at the same O atom). Indeed, an anisotropic Potts model accounts well for the physical behaviour of LaMnO$_3$, rationalising the existence of an orbital order/disorder transition at $T_{\rm{JT}}=750$\,K that couples to a switch from anisotropic to isotropic magnetism and a hundred-fold decrease in resistivity [Figure 4b].\cite{Ahmed_2006} On doping to form the mixed-valence series La$_{1-x}$Ca$_x$MnO$_3$, the fraction of JT-active sites is reduced, the orbital disordered state stabilised, and $T_{\rm{JT}}$ lowered.\cite{Pissas_2005} By $x\simeq\frac{1}{3}$, $T_{\rm{JT}}$ is commensurate with the magnetic ordering temperature ($\simeq300$\,K), and the system exhibits a colossal magnetoresistance (CMR) effect.\cite{Rao_1996} The study of CMR remains an active field, but it is nonetheless clear that the configurational degeneracy of orbital states and the coupling between orbital and magnetic degrees of freedom at these critical temperatures are fundamental ingredients.

\begin{figure}
\begin{center}
\includegraphics{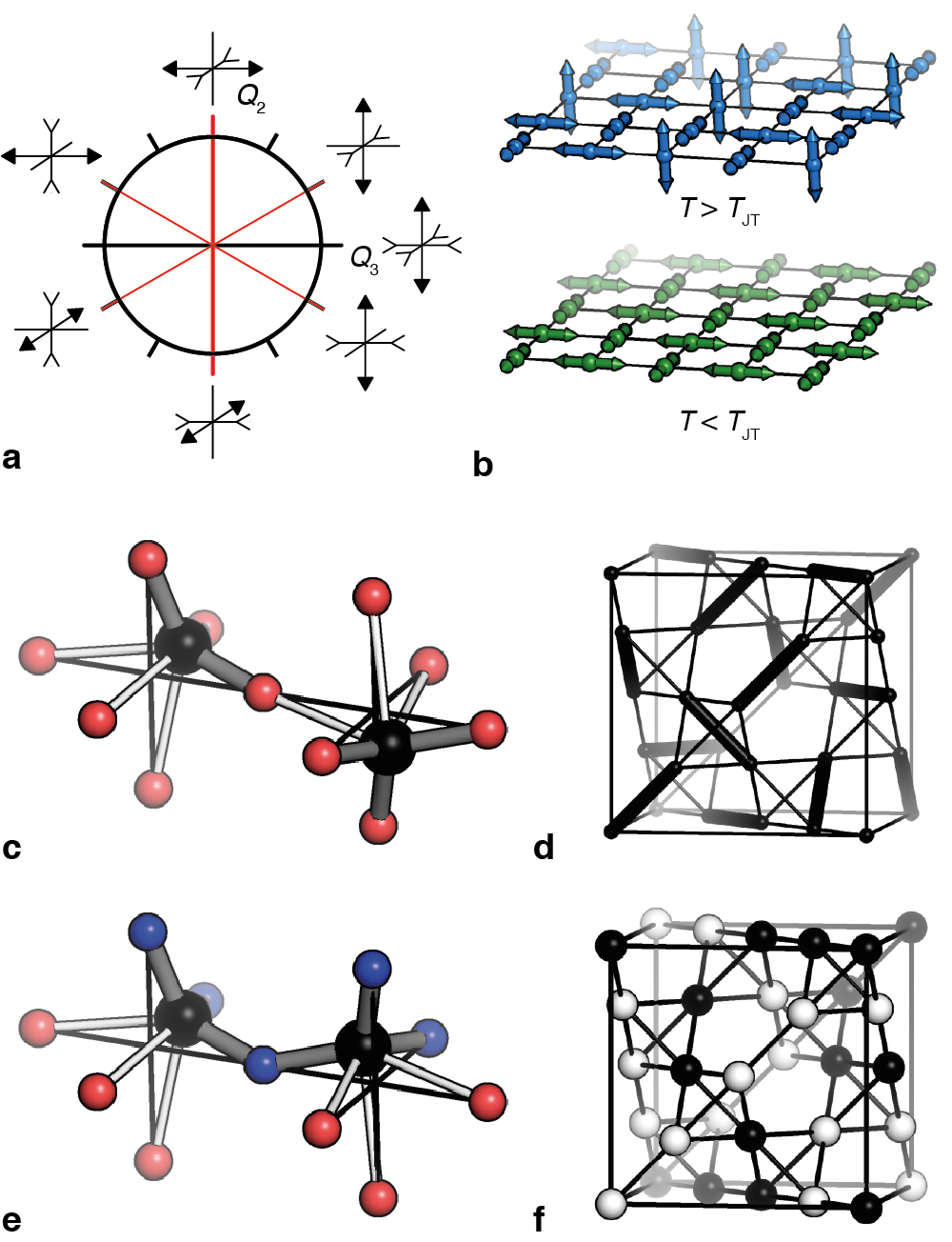}
\end{center}
\caption{\label{fig4} {\bf Correlated disorder from electronic instabilities and compositional complexity.} (a) The $E_g$ distortion space of an octahedrally-coordinated transition-metal ion is spanned by the $Q_2$ and $Q_3$ distortions. In Jahn--Teller active LaMnO$_3$ (Mn$^{3+}$, $d^4$) it is the former type that occurs, giving one pair of long Mn--O bonds, one short pair, and one intermediate pair. There are six equivalent such distortions (red lines). (b)  The orbital order/disorder transition in LaMnO$_3$ involves a progression from alternating $Q_{2,xy}$ distortions at $T<T_{\rm JT}$ to a correlated disordered population of all six $Q_2$ states at $T>T_{\rm JT}$.\cite{Thygesen_2017b} (c) Correlated off-centering of Ti$^{4+}$ ions (black spheres) in BaTiO$_3$ driven by second-order Jahn--Teller effects. The distortions strengthen some Ti--O bonds at the expense of others. That each O atom is bound strongly by just one Ti drives Kitaev-like interactions (Eq.~\eqref{Kitaev}) responsible for complex disorder in BaTiO$_3$.\cite{Comes_1970,Senn_2016} (d) Peierls-type distortions in the hypothetical pyrochlore Y$_2$Nb$_2$O$_7$ would drive a dimer covering of the pyrochlore lattice.\cite{Blaha_2004} (e) Local anion order in EuWO$_{1.5}$N$_{1.5}$ couples W-atom displacements via the same interaction \eqref{Kitaev} as (c) but with the sign of the interaction parameter $J$ inverted. (f) One possible arrangement of Ni (white) and Cr (black) atoms in CsNiCrF$_6$,\cite{Fennell_2019} chosen to emphasise the mapping to the dimer state in (d); note each dimer in the latter is replaced by a pair of cations of the same type in the former.}
\end{figure}

Second-order Jahn Teller (SOJT) distortions can be exploited in a conceptually similar manner. The SOJT effect removes local inversion symmetry so as to allow mixing between orbitals of different symmetries, such as the filled $ns$ and empty $np$ orbitals of the `lone-pair' $p$-block cations (Pb$^{2+}$, Bi$^{3+}$, \emph{etc}.) or the vacant M-$nd$ and filled O-$2p$ states of $d^0$ transition-metal-containing oxides.\cite{Pearson_1975} The direction and magnitude of the displacement responsible for inversion-symmetry breaking acts as a local degree of freedom. As for conventional (first order) JT distortions, the SOJT effect results in the strengthening of some bonding interactions at the expense of others, which again provides a source of coupling between neighbouring sites. In the case of BaTiO$_3$, for example, off-centering of the Ti$^{4+}$ cation along a local $[111]$ diagonal (\emph{i.e.}\ towards one face of the TiO$_6$ coordination polyhedron) strengthens three mutually-perpendicular Ti--O bonds and weakens the other three [Figure 4c].\cite{Comes_1970,Zhong_1995} The tendency for each O atom to participate in one strong and one weak Ti--O bond drives a Kitaev-type anisotropic interaction that couples local displacements into polar chains and is in turn implicated in the emergence of bulk polarisation and ferroelectric response.\cite{Comes_1970,Senn_2016,Wolpert_2018} The same mechanism operates in the ferroelectric KNbO$_3$. Anomalous dielectric (rather than ferroelectric) behaviour is observed in Bi$_2$Ti$_2$O$_7$, for which Bi$^{3+}$ cations are arranged on the pyrochlore lattice.\cite{Melot_2009} Collective inversion-symmetry breaking at the Bi$^{3+}$ site now results in a geometrically frustrated state conceptually related to that of the spin-ices. It is the shallow configurational landscape of this phase that allows its anomalously large dielectric response to applied field.

Metal--insulator transitions (MITs) of the Peierls type also arise from JT-like instabilities. The one-dimensional nature of $d^1$ MO$_2$ phases (M = V or Nb) leads to long-range symmetry breaking at the MIT as chains form of alternating short M--M dimers and long M$\ldots$M pairs.\cite{Morin_1959,Attfield_2015} By contrast, the analogous transition would be hidden for Y$_2$Nb$_2$O$_7$ as there is a large configurational degeneracy of Nb--Nb dimer decorations on its pyrochlore lattice [Figure 4d].\cite{Blaha_2004,McQueen_2008} In the related $d^2$ system Y$_2$Mo$_2$O$_7$, the same type of orbital-ice phase results in a large variance in magnetic superexchange strengths that in turn stabilises an unusual spin-glass ground state.\cite{Shinaoka_2013,Thygesen_2017} These systems are specific examples of disordered `orbital molecule' phases---semiconducting states in which the unpaired electrons of transition-metal cations are either shared to form metal--metal bonds or delocalised across a handful of neighbouring sites (\emph{e.g.} the trimerons of Fe$_3$O$_4$).\cite{Attfield_2015,Senn_2012} It increasingly seems that many orbital-molecule phases exhibit $n$-mer order/disorder transitions, such as that involving trimer/tetramer arrangements in AlV$_2$O$_4$,\cite{Browne_2017} fluctuating dimers in CuIr$_2$S$_4$,\cite{Bozin_2019} and the formation of a trimeron liquid in Fe$_3$O$_4$.\cite{Perversi_2019} Just as JT disordering in LaMnO$_3$ has a profound effect on electronic properties of the system, one might anticipate similarly important electronic anomalies associated with orbital-molecule disorder. Indeed there are strong conceptual parallels to complex ordering in zintl or intermetallic phases\cite{Folkers_2018} and also to the resonance valence bond model originally proposed by Anderson in the context of high-temperature superconductivity.\cite{Anderson_1973}

So, from a design perspective, transition-metals and/or $p$-block cations with particular electronic configurations can be chosen to favour very specific types of local degrees of freedom. JT-active first-row transition-metal cations will exhibit collective JT phases with antiferro-type nearest-neighbour interactions; SOJT-active cations will give XY or Heisenberg degrees of freedom that will couple \emph{via} ferroic Kitaev-type interactions (although dipolar contributions may also be important); and $4d/5d$ transition metals with open shell configurations may form `hidden' gapped phases in which cations have displaced in an antiferrodistortive sense. In each case, the nature of the underlying lattice is crucial in stabilising the relevant disordered state. 





\noindent{\bf Compositional complexity.} To the chemist, a very natural mechanism by which disorder might be introduced into a crystal structure is to employ different components that are capable of occupying the same crystallographic site. Alloys are an obvious example; so too are the many solid solutions routinely studied throughout materials chemistry. In many cases, compositional disorder is essentially random; such is the basis, after all, of the paradigms of Vegard's law and the tuning of `chemical pressure' (to take two examples). The less trivial cases are those in which compositional disorder is strongly correlated, such as occurs whenever local rules are important and/or geometric frustration plays a role.\cite{Castellanos_1980} In the discussion that follows, we include vacancies as a kind of compositional variable, although there will be cases (precisely as discussed in Ref.~\citenum{Parsonage_1978}) where the distinction between vacancy and displacive (dis)order is semantic rather than physical.

Certainly a very clear example of nontrivial compositional disorder occurs in oxynitride perovskites such as SrMO$_2$N (M = Nb, Ta).\cite{Yang_2011,Camp_2012} The octahedral coordination environment of each M$^{5+}$ cation consists exclusively of four O and two N atoms, with the two nitride ligands arranged \emph{cis} to one another in order to allow $\pi$-donation into orthogonal $t_{2g}$ orbitals. Despite this prescribed coordination environment, the system is able to maximise its configurational entropy by adopting a disordered arrangement containing zig-zag M--N--M linkages (the `S2C' phase of Ref.~\citenum{Overy_2016}, and an example of a spaghetti phase). Depending on the precise chemistry and synthesis temperature it is possible to navigate different phases in which this zig-zag disorder extends in either two or three dimensions.\cite{Johnston_2018} In all cases, the point symmetry of the MO$_4$N$_2$ octahedron is effectively $C_{2v}$. This allows the central M cation to displace towards the edge spanned by both nitrides (\emph{i.e.} the local $C_2$ axis), giving rise to a local dipole. Hence this compositional decoration drives a collective off-centering that is conceptually related to that arising from SOJT effects as described above [Figure 4c]. As for the SOJT systems, the off-centre displacements are correlated by a Kitaev-like interaction of the form
\begin{equation}
\mathcal{H} = J\sum_{\mathbf r}\left(e_{\mathbf r}^xe_{\mathbf r+\mathbf a}^x + e_{\mathbf r}^ye_{\mathbf r+\mathbf b}^y + e_{\mathbf r}^ze_{\mathbf r+\mathbf c}^z\right),\label{Kitaev}
\end{equation}
where superscript $x,y,z$ denote the Cartesian components of the displacement vector $\mathbf e_{\mathbf r}\in\{110\}$ at lattice site $\mathbf r$, and $\mathbf a,\mathbf b,\mathbf c$ are the lattice vectors.\cite{Wolpert_2018} What differs between the two cases is the sign of the coupling constant $J$ and hence the type of local order that emerges: in the oxynitrides, $J>0$ and displacements tend to alternate in direction from site to site, whereas collective polarisation emerges in the SOJT systems because $J<0$. Moreover, the off-centre displacements are effectively fixed during synthesis for the former, but may be driven between equivalent states in the latter. An ongoing challenge in this area is to establish a clear link between this very prescribed type of compositional disorder and the various attractive physical properties of oxynitrides, such as their high dielectric constants, photocatalysis, and CMR behaviour.\cite{Yang_2010}

Entirely analogous correlated compositional disorder may be expected in many other important mixed-anion phases, not least the oxyfluorides and oxyhydrides.\cite{Goto_2017,Kageyama_2018} In all cases, bonding considerations provide a strong set of local rules that---depending on the exact chemistry, synthesis conditions, and lattice geometry---need not drive long-range order.

When two components are substituted on the same crystallographic site, composition behaves as an Ising variable, and hence any complex disordered state based on Ising degrees of freedom might in principle be realised through judicious choice of the composition and underlying crystal geometry. For example, the two-in-two-out ice states of the Ising pyrochlore antiferromagnet are realised \emph{via} the the distribution of Cr$^{3+}$ and (non-magnetic) Sb$^{5+}$ in RE$_2$CrSbO$_7$ pyrochlores (RE = Y, Yb, Dy, Er).\cite{Whitaker_2014} Here the ambient-temperature magnetic behaviour is dominated by ferromagnetic interactions between Cr$^{3+}$ ions, which are unfrustrated despite the geometrically frustrated cation ordering. Not so for the defect pyrochlore CsNiCrF$_6$, where there is strong interplay between the icelike arrangement of Ni$^{2+}$ and Cr$^{3+}$ and the now-frustrated Heisenberg antiferromagnetic interactions between those ions [Figure 4f].\cite{Fennell_2019} Similar effects are seen in RbFe$_2$F$_6$, for which charge order of Fe$^{2+}$/Fe$^{3+}$ (of the type first proposed by Anderson for Fe$_3$O$_4$, Ref.~\cite{Anderson_1956}) provides the underlying compositional complexity.\cite{Kim_2012} These systems are rare examples of multiple Coulomb phases because they each support two emergent gauge fields---one involving composition (or charge) and the other magnetism. The same concept extends naturally beyond the pyrochlore lattice: YbMgGaO$_4$ is a classical spin liquid stabilised by compositional disorder of Yb and Mg on the triangular lattice.\cite{Zhang_2018} The use of correlated compositional disorder to drive unconventional magnetism is an emerging frontier of condensed-matter physics.

In a more obviously chemical context, non-random cation disorder in rocksalt oxides is increasingly viewed as a design parameter in the application of these systems as cathode materials.\cite{Lee_2014,Ji_2019} Taking as the general case the composition LiMO$_2$, where M is a trivalent transition-metal cation (or mix of cations with average charge 3+), the tendency to distribute the more highly charged M$^{3+}$ species so as to avoid M$\ldots$M neighbours gives rise to an effective antiferroic interaction between compositional Ising states. These are frustrated on the fcc Li/M sublattice of the rocksalt structure, which no doubt accounts the complexity of this general family. Hence Li$^+$-ion arrangements are driven by longer-range interactions, which may or may not enforce long-range order. Order facilitates structural characterisation, but it also results in anisotropy---such as in layered LiCoO$_2$---that increases strain during charge/discharge cycles and hence can reduce battery lifetimes.\cite{Yabuuchi_2016} There is now intense interest in the manipulation of short-range order within isotropic disordered systems to optimise battery performance.


The competition between composition and lattice geometry can be exploited to favour disordered states. For example, antiferroic interactions are not frustrated on the simple cubic lattice. In our compositional mapping, this accounts for the well-known family of double-perovskites, for which B-site cations of one type are connected exclusively to B-site cations of the other type. However, complexity can be introduced in these systems by varying the ratio of substituents from 1:1 to 1:2. This is precisely the composition that optimises the complexity of domain structure in relaxor ferroelectrics (\emph{e.g.}\ PbMg$_{1/3}$Nb$_{2/3}$O$_3$) and relaxor ferromagnets (\emph{e.g.}\ LaNi$_{2/3}$Sb$_{1/3}$O$_3$).\cite{Pasciak_2012,Battle_2013} The tendency for like B-site cations to avoid occupying neighbouring sites automatically partitions the underlying cubic lattice into its two constituent interpenetrating fcc lattices: one of which is occupied exclusively by the majority B-site cation, and the other occupied by both types in a 1:2 ratio (again frustrated).\cite{Bostrom_2019} Hence, Prussian Blue analogues of general formula M$^{\rm{II}}$[M$^{\rm{III}}$(CN)$_6$]$_{2/3}\Box_{1/3}$ ($\Box$ = vacancy) are also of this same structure type and show a similar degree of structural complexity that now affects the distribution and connectivity of vacancies.\cite{Simonov_2019} In this particular family, as in the disordered rocksalt cathode materials discussed above, there are now signs that the type of local order present in the disordered state might be tuned rationally by varying chemical composition and/or synthesis approach---an appealing kind of defect engineering.\cite{Simonov_2019} Similar approaches have been used to optimise superconducting transition temperatures in YBa$_2$Cu$_3$O$_{6+x}$ crystals,\cite{Veal_1990} and are currently being explored in the field of MOFs,\cite{Fang_2015} where the concentration---if not yet arrangement---of vacancy clusters (\emph{e.g.}\ in UiO-66 and its derivatives) can be used to tune mechanical, sorption, and catalytic behaviour.\cite{Shearer_2016,Cliffe_2015}


For each of these examples, compositional disorder is imprinted during synthesis and hence is presumed static. But in certain systems with large vacancy concentrations and/or suitably polarisable ions, it is possible for compositionally disordered states to respond dynamically to external stimuli. A number of extremely well known classes of disordered crystals fall into this category. One is that of the fast-ion conductors. To take the paradigmatic example of $\alpha$-AgI, the Ag$^+$ cations are distributed on a multi-site sublattice that is dominated by vacancies.\cite{Nield_1993} The hopping barrier is sufficiently low that at readily-accessible temperatures the Ag$^+$ sublattice behaves as if liquid. In the oxide-ion analogues, such as Bi$_2$O$_3$ and yttria-stabilised zirconia, this ionic mobility is widely exploited in fuel cell technology.\cite{Wachsman_2011} A second examplar of responsive occupational disorder is given by the many proton-based ferroelectrics, such as KH$_2$PO$_4$ (KDP).\cite{Grindlay_1959} In these systems, H$^+$ ions can occupy one of two equivalent (or nearly equivalent) sites, and the application of an external field can induce flipping from one site to the other with a concomitant change in bulk polarisation. In proton or lithium-ion conductors, the available sites for H$^+$/Li$^+$ ions form connected paths to allow bulk transport in the presence of an ion gradient; the same design rules of site degeneracy and low inter-site barrier heights apply.\cite{DiStefano_2019}


\noindent{\bf Directed self-assembly.} The nature of proton disorder in water ice, although couchable in terms of hydrogen-bond (Ising) degrees of freedom, arises spontaneously from the shape and charge distribution of the H$_2$O molecule itself. In this sense one might speak of its 2-in-2-out disorder as `encoded' within the molecule, and the emergence of a strongly-correlated disordered state as a process of directed self-assembly.\cite{Pauling_1935,Bernal_1933} Since the four hydrogen bonds made by each H$_2$O molecule are tetrahedrally disposed, the vast majority of solid ice phases are based on tetrahedral nets.\cite{Salzmann_2011} For most of these, the number of different decorations by H$_2$O molecules satisfying sensible hydrogen-bonding rules actually scales with system size. Hence the configurational entropy associated with orientational disorder is extensive, in turn stabilising the disordered state.\cite{Parsonage_1978} The dielectric properties of ordered and disordered ices are fundamentally different, and indeed many ordered phases are polar.\cite{Bramwell_1999} But perhaps the most important physical implication of orientational disorder in water ice is its effect on molar volume: disorder expands the hydrogen-bonded network to the extent that ordered ice would sink rather than float in water.\cite{Salzmann_2011}




The concept that molecular components of appropriate design might encode for complex disordered states is illustrated neatly by a number of 2D (or quasi-2D) supramolecular assembles. The approach taken in Ref.~\citenum{Blunt_2008} develops from the observation that trimesic acid self-assembles into a hydrogen-bonded honeycomb when deposited onto an appropriate surface.\cite{Barth_2005} The larger molecule {\it p}-terphenyl-3,5,3$^\prime$,5$^\prime$-tetracarboxylic acid (TPTC) is a steric and functional mimic of a trimesic acid dimer pair, where the intra-dimer hydrogen bonds are replaced by covalent bonds. So TPTC also self-assembles into a honeycomb, but with one third of the links---exactly one per node---now covalent [Figure 5a].\cite{Blunt_2008} The presence of correlated disorder leads to an unexpectedly dynamic assembly, in which defects behave as emergent topological charges.\cite{Stannard_2012} A conceptually similar mesoscopic analogue forms spontaneously from the aggregation of dumbells (connected circular pairs): the aggregate resembles a triangular packing of circles, but each is a member of exactly one dimer.\cite{Gerbode_2010,Muangnapoh_2014} Both examples reflect a common design strategy of engineering dimer decorations of ordered lattices.\cite{HajiAkbari_2015} Since the problem of distributing dimers on the honeycomb lattice (perhaps unexpectedly) maps directly onto the TIA, TPTC assemblies have exactly the same configurational entropy and excitations as that model system.\cite{Overy_2016} In this case, the corresponding Ising degree of freedom, effective antiferroic interactions, and underlying triangular lattice are emergent properties of the TPTC assembly. Such mappings allow indirect routes for the design of specific phases.

\begin{figure}
\begin{center}
\includegraphics{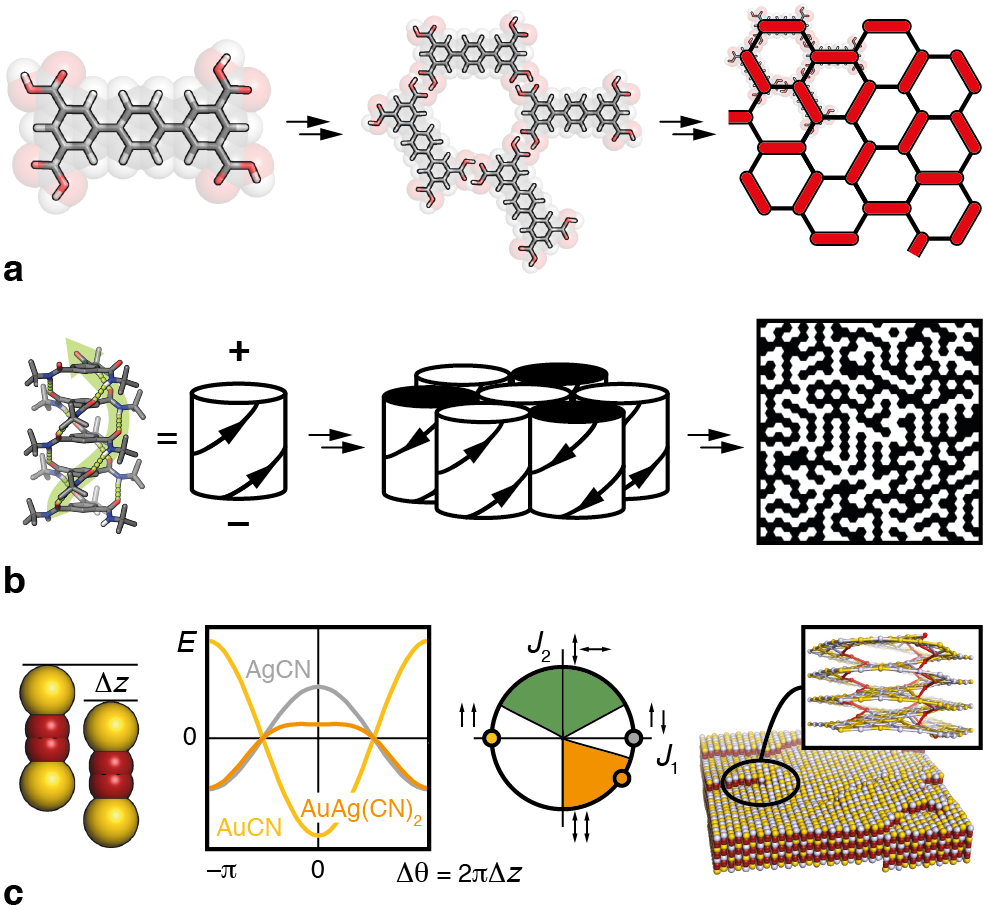}
\end{center}
\caption{\label{fig5} {\bf Correlated disorder by self-assembly.} (a) TPTC molecules assemble in 2D through formation of hydrogen-bonded carboxylate dimers to give a disordered network describable in terms of a dimer decoration of the hexagonal lattice.\cite{Blunt_2008} (b) Directional hydrogen-bonding interactions (green arrow) give a bulk polarisation to columns of BTA molecules. Dense assemblies of these columns involve triangular packing, with adjacent stacks tending to adopt inverse polarisations (black and white, here). The corresponding state relates to the triangular Ising antiferromagnet.\cite{Simonov_2014,Zehe_2017} (c) Self-assembly of the columnar cyanides Au$_{1-x}$Ag$_x$CN is driven by inter-chain interactions, which vary periodically with relative chain displacement $\Delta z$, and hence effective phase $\theta=2\pi z$. These interactions take the form $E=J_1\cos(\Delta\theta)+J_2\cos^2(\Delta\theta)$, which relates the phase behaviour of the family to that of the triangular bilinear--biquadratic XY magnets. Topological charges (vortices/antivortices) in these magnets are mapped onto screw dislocations in the cyanides.\cite{Cairns_2016}}
\end{figure}

TIA assemblies can also be generated through self-assembly in much more direct means. By way of example, substituted 1,3,5-benzenetrisamides (BTAs) are known to assemble spontaneously into columnar phases.\cite{Miyajima_2012,Fitie_2010} The approximately-planar molecules stack through $\pi$-$\pi$ interactions into columns, that are further stabilised by helical 1D strands of amide hydrogen bonds on their edges. This hydrogen bonding is strongly directional and gives each molecular stack an electric dipole moment oriented parallel to the stacking axis. The up/down sense of this dipole is an Ising degree of freedom.\cite{Simonov_2014} In the solid, columns form a dense packing to give a triangular arrangement of Ising states. Dipole--dipole interactions between neighbouring columns provide the final ingredient of antiferroic coupling, such that the TIA phase emerges spontaneously [Figure 5b].\cite{Simonov_2014,Zehe_2017} From a magnetic viewpoint, similar design rules stabilise the TIA in the magnetocaloric Tb(HCOO)$_3$:\cite{Harcombe_2016} Tb$^{3+}$ ions form ferromagnetic chains that are arranged on a triangular lattice and are coupled antiferromagnetically by superexchange across formate bridges. Returning to the trisamide family, there is a particular functional implication of the configurational degeneracy of the TIA state: it renders the system especially sensitive to weak longer-range interactions, which can be tuned by varying substituents on the trisamide core.\cite{Zehe_2017} The system can be optimised to stabilise polar nanodomains, a motif that is thought to enhance their functional role as crystallisation nucleation agents.

A particularly appealing strategy for controlling the emergent lattice geometry comes from the domain of liquid crystal science. It was shown in Ref.~\citenum{Zeng_2011} that `tetraphilic' X-shaped molecules assemble into various columnar phases, the geometries of which are determined by the relative volumes and lengths of their different substituents. In the assemblies that form, each X-shaped molecule uses two of its four structural components to form the columnar structure; each of the other two extend laterally into each of two neighbouring channels (one on either side of the column wall). If these lateral components are chemically distinguishable, then there are two possible orientations and again we have an Ising system, now on a lattice predetermined by steric considerations. Ref.~\citenum{Zeng_2011} introduces effective antiferroic interactions by making one lateral component hydrophilic and the other hydrophobic, and demonstrates explicitly the formation of a kagome ice phase. Here, the key outcome is probably one of illuminating design strategy rather than any particular functional response.

In two dimensions, XY models tend to offer a richer physics than do their Ising analogues: a well-known example is the formation of vortex/antivortex pairs that obey their own emergent physics and might be used in data storage.\cite{Kosterlitz_1973} The vertical position $z$ of periodic chains behaves as an effective XY degree of freedom \emph{via} the simple mapping $\theta=2\pi z$.\cite{Cairns_2016} Displace a chain along its axis by one repeat length and the system is unchanged, in the same way that rotation by $2\pi$ maps a planar spin back onto itself. Consequently, assemblies of periodic chains can adopt the same phase behaviour as 2D XY models. Probably the best understood example is that of the linear cyanide polymers Ag$_{1-x}$Au$_x$CN, all of which involve triangular packings of periodic chains.\cite{Cairns_2016} The effective pairwise interactions are governed by competition between metallophilic and electrostatic contributions; the balance between the two varies as a function of composition $x$. For AgCN ($x=0$), electrostatics win out. Chain neighbours are best stabilised by relative shifts $\Delta z=\frac{1}{2}$ that place the CN$^-$ anions of one chain next to the Ag$^+$ cations of the other. Hence the interaction between effective XY degrees of freedom is antiferroic. The situation reverses for AuCN: aurophilic interactions more strongly favour coalignment of neighbouring chains and the effective XY interaction is ferroic. At intermediate compositions, such as Ag$_{1/2}$Au$_{1/2}$CN, the energetics are extremely finely balanced, and higher-order (biquadratic) interactions stabilise a frustrated nematic phase that contains a high density of vortex/antivortex pairs [Figure 5c].\cite{Zukovic_2003} So the family provides an attractive example of using composition to navigate phase space that might be exploited to target specific states of particular interest. In principle, the phase behaviour of the diverse families of urea and thiourea clathrates may yet be interpretable (and manipulatable) in this same context.\cite{Schlenk_1950} 

Just as 2D physics can be realised in columnar phases, so is it that 1D physics emerges in layered systems. Here, as ever, Ising models are the best studied.\cite{Yeomans_1988} The canonical example is that of close-packing, where successive layers can be in one of two equivalent arrangements, and cubic and hexagonal close packing arise from effective ferroic and antiferroic interactions, respectively, between next-neighbour layers. The structures of stacking-faulted ices have been rationalised in this context by establishing that the coupling strength is comparable to the available thermal energy at ice formation.\cite{Playford_2018} A similar effect is likely responsible for stacking disorder in Ni(CN)$_2$, despite the difference in layer geometry.\cite{Goodwin_2009b} By contrast, the complex polymorphism of silicon carbides is understood in terms of competing interactions between successive layers that map their phase behaviour onto the ANNNI model.\cite{Yeomans_1988} One of the key targets of this area is to use complex layering motifs to reduce thermal conductivity, \emph{e.g.}\ in layered chalcogenides.\cite{Jood_2015} As in the columnar cyanides, layer composition appears to be the most straightforward---if still empirical---design variable for navigating the effective 1D phase space spanned by these systems.

\noindent{\bf Molecular shape as a symmetry lowering mechanism.} An important theme of soft-matter science is that shape alone can drive extraordinary phase complexity. This complexity arises from two effects: a shaped object has additional orientational degrees of freedom that a spherical particle does not; and the interactions between shaped particles are inherently anisotropic. In a survey of the dense packings of 145 high-symmetry convex polyhedra, only 24 were found to form ordered crystalline phases (including the remarkable quasicrystalline tetrahedra packing).\cite{Damasceno_2012,Engel_2015} Polyhedra with large anisotropy favour liquid crystalline states; those with low anisotropy tend to form plastic crystals instead. The inverse relationships have long been used by chemists as design principles: hexahydroxytriphenylene derivatives make good discotic liquid crystals,\cite{Kumar_2000} where there is strong orientational order but weak positional order; C$_{60}$ is easily stabilised in a plastic phase, where the molecular centres of mass order but orientations do not.\cite{David_1993} In these soft materials, one relies on supramolecular interactions to determine the geometry of the lattice, which makes it challenging to anticipate phase behaviour \emph{a priori}. Nevertheless a number of important paradigmatic systems, from simple solids such as CO$_{(\rm s)}$ through to the many industrially-important families of liquid crystals fall into precisely this category.

Ref.~\citenum{Parsonage_1978} distinguishes between the role of molecular shape in these molecular crystals and its role in ionic molecular salts. In the latter case, electrostatic interactions help organise molecular units onto a specific lattice, and the interactions between molecules are no longer direct but are longer-range and/or mediated by the non-molecular component, whatever it may be. Hence there is a more obvious strategy for engineering disordered systems of a specific type. The difference in point symmetry between the molecule (local) and its crystallographic site (average) effectively determines the molecular orientational degrees of freedom. The underlying lattice is of course constrained by the crystal structure. And the effective interactions between neighbouring molecules are often dominated by dipolar terms and/or coupling to strain. From a statistical mechanical viewpoint, the alkali cyanides (\emph{e.g.}\ KCN) and their derivatives are probably the best known examples;\cite{Pauling_1930,Lewis_1986} molecular ferroelectrics such as NaNO$_2$ have also long been studied.\cite{Yamada_1963}

Our focus here is on systems of more recent interest, where molecules are contained within the cavities of a host framework. The examples of greatest currency are certainly the organic/inorganic hybrid perovskites,\cite{Li_2017} but metal--organic frameworks and clathrates are also of this type. From a design perspective, the key attraction is that the geometry of the host strictly determines the geometry of the lattice on which the molecular components are arranged. The strategy of exploiting differences in local and crystallographic symmetry to control microscopic degrees of freedom applies equally well; orientational coupling between neighbouring molecules is often dominated by dipolar and/or higher-order multipolar interactions. Moreover, there is particular scope for interplay between orientational disorder and materials properties \emph{via} coupling to the host framework.

A simple example of the phenomenology of order/disorder transitions in hybrid perovskites is given by the guanidinium cyanoelpasolites.\cite{Coates_2019} The guanidinium cation has a large quadrupole moment---it is essentially flat. Importantly, its molecular point symmetry ($D_{3h}$) is not a supergroup of the point symmetry ($T_d$) of the A site in the aristotypic elpasolite (double perovskite) lattice. There are four equivalent ways of lowering the local symmetry appropriately that correspond to alignment of the guanidinium quadrupole along the four $\langle111\rangle$ cube diagonals. Hence guanidinium orientations behave as 4-state Potts degrees of freedom. In these particular materials, coupling to strain gives effective ferroquadrupolar interactions ($|J|\simeq100$\,K), the strength of which can be tuned through chemical pressure of the surrounding lattice: increasing lattice size acts to decrease $|J|$. A quadrupolar order/disorder transition occurs at $T_{\rm c}\simeq4.7J$ that follows closely the simple 4-state Potts ferroquadrupolar model on the simple cubic lattice. This scale of $T_{\rm c}$, which is representative of orientational order/disorder transitions in molecular perovskites, is clearly attractive for device applications. Entirely analogous phase behaviour drives the dielectric anomalies in molecular formate perovskites with polar A-site cations (in these, the interactions are nearly always antiferrodipolar) and indeed the ferroelectric/paraelectric transitions in the topical all-organic perovskites [MDABCO][NH$_4$]X$_3$ (MDABCO = N-methyldabconium; X = Cl, Br, I).\cite{Ye_2018} Simple Ising or Potts-type models seem to account well for the experimental phase behaviour of those hybrid perovskites to which they have been applied.\cite{Simenas_2016,Coates_2019,Simenas_2019}

The entropy change associated with these orientational order/disorder transitions can be exploited in the design of barocalorics for solid-state cooling [Figure 6a].\cite{Lloveras_2019} The key example in this context is [TPrA][Mn(dca)$_3$] (TPrA$^+$ = tetrapropylammonium; dca$^-$ = dicyanamide).\cite{BermudezGarcia_2017} At ambient pressure, the system exhibits an order/disorder transition at $T_{\rm c}=330$\,K involving internal conformational degrees of freedom of the TPrA$^+$ cation. The entropy change involved in this transition $\Delta S=18.7$\,J\,K$^{-1}$\,mol$^{-1}$ is amplified by the coupling between TPrA$^+$ conformations and orientations of the dca$^-$ framework linkers. On application of hydrostatic pressure, $T_{\rm c}$ increases rapidly ($\partial T_{\rm c}/\partial p=23.1$\,K\,kbar$^{-1}$),forming the basis of a barocaloric cooling strategy [Fig. 6 a--c]. There is a very natural parallel to magnetocalorics, for which ordering is driven by an external magnetic field rather than hydrostatic pressure. Frustrated magnets such as Gd$_3$Ga$_5$O$_{12}$ and Tb(HCOO)$_3$ tend to form particularly good magnetocalorics:\cite{Tishin_2003,Lorusso_2013,Saines_2015} their shallow configurational landscape imparts a heightened susceptibility to external field that in turn reduces the ciritical field $B_{\rm c}$ at which ordering occurs. The same strategy of exploiting geometric frustration might in principle allow optimisation of barocaloric response in analogues of [TPrA][Mn(dca)$_3$].

\begin{figure}
\begin{center}
\includegraphics[width=8.3cm]{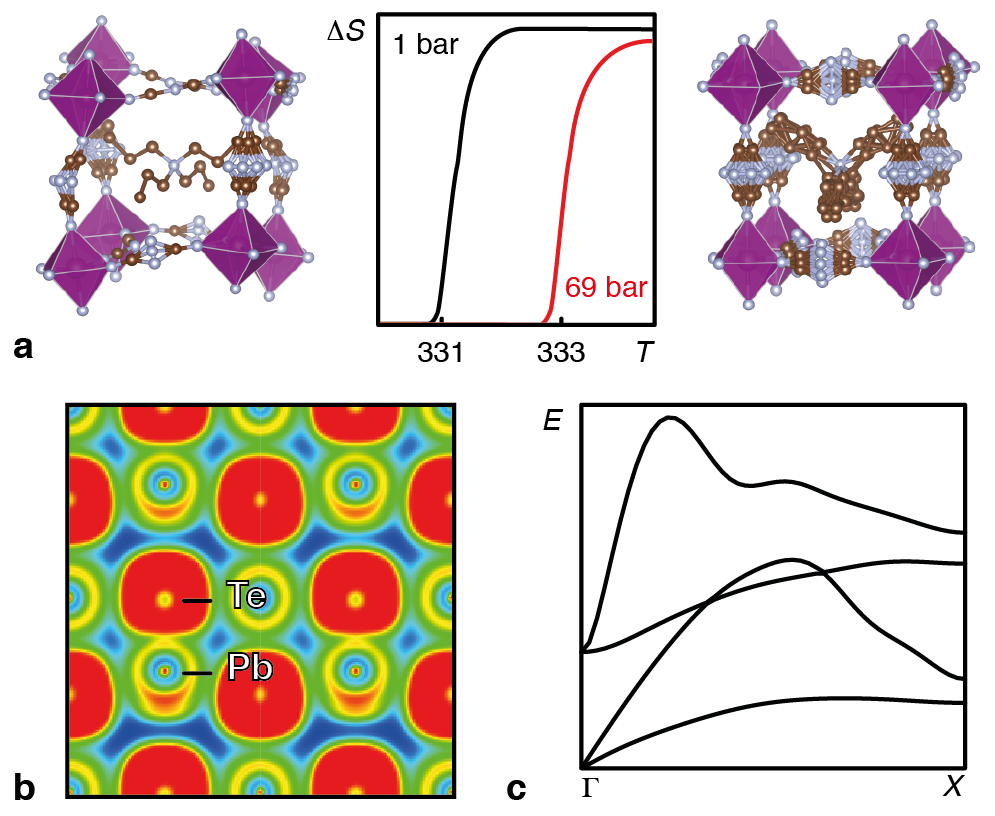}
\end{center}
\caption{\label{fig6} {\bf Low-energy dynamics.} (a) The hybrid perovskite [TPrA][Mn(dca)$_3$] exhibits an conformational order--disorder transition on cooling below 331\,K. The ordered state can be stabilised by applying pressure, which provides a mechanism for barocaloric cooling.\cite{BermudezGarcia_2017} (b) A representation of the valence charge distribution in PbTe upon activation of the low-energy optic phonons; note the emergence of a lone-pair on the Pb$^{2+}$ cation.\cite{Sangiorgio_2018} (c) This emergent SOJT distortion results in an amalously low energy of the zone-centre optic branch and softening of the transverse acoustic branch of the phonon dispersion. Both features are implicated in the favourable themoelectric behaviour of PbTe.\cite{Sangiorgio_2018}} 
\end{figure}

Orientational disorder of the MA$^+$ cations is thought to play an important role in preventing exciton recombination in the photovoltaic MAPbI$_3$.\cite{Frost_2014,Herz_2018} MA$^+$ reorientations provide a strongly varying dipolar field at the perovskite A-site, which in turn helps screen the Coulombic electron--hole interaction. At the same time, molecular degrees of freedom couple to large-scale distortions of the surrounding [PbI$_3$]$^-$ lattice that help localise electronic states. The distortions are amplified by the susceptibility of the Pb$^{2+}$ cation to SOJT effects, which may couple in a manner related to that discussed above for ferroelectric BaTiO$_3$/KNbO$_3$.\cite{Laurita_2017} Hence the lead halide perovskites possess a large number of entangled degrees of freedom such that the timescales associated with various ordinarily-separate processes are now linked: electron localisation, dipolar field fluctuations, cation off-centering, lattice distortions, and MA$^+$ molecular reorientations. In terms of design rules, the role of shape is to break A-site symmetry over timescales longer than that of uninhibited exciton recombination, and the role of disorder to avoid collective structural and electronic relaxation.

Polarisability is a kind of incipient shape. CsPbI$_3$ is an effective analogue of MAPbI$_3$ because the spherical electron density of the large and polarisable Cs$^+$ ion is easily perturbed by surrounding fields.\cite{Zhu_2019} So the low-energy distortions of the anionic PbI$_3^-$ framework can easily couple to charge redistribution at the Cs$^+$ site in the same way they might couple to MA$^+$ reorientations in MAPbI$_3$.
 

\noindent{\bf Low energy dynamics: incipient disorder.} In these dynamically disordered systems, the configurational landscape of related states is sufficiently shallow to be navigated thermally. Hence the design strategies that drive correlated (static) disorder can also give rise to anomalous low-energy dynamics, whereby configurational degeneracy is mapped onto a dispersionless manifold of low-energy excitations (in the magnetism and electronics literature these are often termed `flat bands' \cite{Leykam_2018}) by reducing the barrier height for state interconversion. A good example is that of the rocksalt thermoelectrics such as PbTe.\cite{Delaire_2011} The susceptibility of the Pb$^{2+}$ $6s^26p^0$ configuration to a SOJT distortion favours local symmetry lowering that couples \emph{via} the same Kitaev-like interactions operating in BaTiO$_3$. Whereas in PbO the distortion is static, in PbTe it is dynamic. This softens the optic phonons, giving a shallow low-energy phonon dispersion that drives the reduced thermal conductivity exploited in its thermoelectric behaviour [Figure 6b,c].\cite{Sangiorgio_2018} An entirely analogous mechanism drives the anomalously large dielectric response of Bi$_2$Ru$_2$O$_7$, for which dynamic off-centering of Bi$^{3+}$ ($5s^25p^0$) is now key.\cite{Avdeev_2002} In zeolites, there are so many different mechanisms by which correlated rotations and translations of Si/AlO$_4$ tetrahedra can occur and the barriers between different displacement patterns so low that dispersionless bands of rigid-unit mode (RUM) excitations are dragged into the low-energy phonon spectrum.\cite{Mukhopadhyay_2004} An important implication of the density of these RUMs in $k$-space is the ability of zeolites to support localised distortions with very low energy penalties.\cite{Hammonds_1997} This kind of local flexibility mimics the conformational flexibility of enzyme active sites and is thought to favour catalytic efficiency and ion transport in zeolites.\cite{Sartbaeva_2012}



An interesting balance between static and dynamic correlated disorder occurs in the cubic negative thermal expansion (NTE) material ZrW$_2$O$_8$.\cite{Mary_1996} The dominant incipient structural distortion in this system involves concerted translations and rotations of WO$_4$ tetrahedra that act to connect these units into one-dimensional chains (`spaghetti'), at once both increasing the W coordination number from four to five and reducing the system volume.\cite{Keen_2007,Baise_2018} The crystal symmetry is such that there is no unique way of forming these chains, and a simple Pauling-type calculation suggests a configurational entropy of about $R\ln(9/8)$ if the spaghetti-forming distortions were static.\cite{Baise_2018} Because this configurational entropy is extensive, the corresponding dynamic fluctuations must be dense in $k$-space, which is why there are so many volume-reducing phonon modes and hence why ZrW$_2$O$_8$ shows NTE. Under applied pressure or (somewhat bizarrely) on hydration, the same set of distortions are activated in a static sense (\emph{i.e.}\ the interconversion barrier height raised); this gives pressure-induced amorphisation (PIA) in the former case and negative hydration expansion (NHE) in the latter. Hence a single common landscape of correlated disordered states is responsible for each of the unusual phenomena of NTE,\cite{Mary_1996} NHE,\cite{Duan_1999} and PIA\cite{Perottoni_1998} in this one material.

\section{Future directions}

{\bf Adaptive materials from disordered states.} A common theme amongst the many applications of disordered crystals is the importance of configurational degeneracy and the heightened susceptibility of disordered states to perturbation by external stimuli.\cite{Moessner_2006} In these respects of degeneracy and susceptibility there is a strong conceptual parallel to the axioms of dynamic combinatorial chemistry.\cite{Otto_2002} A combinatorial system is highly degenerate because one library can in principle sample a very large number of discrete molecular entities; it is susceptible because host--guest interactions readily perturb the system equilibrium to bias the configurational landscape towards a smaller subset of now-favoured states. Dynamic combinatorial chemistry exploits the configurational flexibility of a suitable library to identify optimised receptors for a given guest. Might not one try to apply similar concepts in the context of disordered crystals? The ability for certain disordered crystals to navigate reversibly their relevant configurational space (\emph{e.g.}\ through molecular reorientations, spin flips, hydrogen-bond inversion$\ldots$) reflects the same kind of configurational flexibility found in combinatorial libraries, distinct from that of the conventional mechanical flexibility of \emph{e.g.}\ wine-rack frameworks. By this mechanism, a single disordered crystal might in principle be able to adapt in different ways to optimise its interactions with a variety of different adsorbates, for example.\cite{Serre_2007} Or to accommodate the deformations induced by a variety of different strains. Or to heal itself after suffering some form of internal damage.

{\bf Correlated disorder engineering.} The unusual gauge field symmetries of certain disordered crystals also offer all sorts of potential avenues for further investigation. For example, one might use this type of symmetry as a key element in the design of hybrid improper ferroelectrics \emph{i.e.}\ to avoid reliance on lone-pair-active cations such as Pb$^{2+}$.\cite{Wolpert_2018} Alternatively, the coexistence of short- and long-range correlations (\emph{e.g.}\ as typified by the `pinch-point' magnetic scattering features of spin-ices\cite{Fennell_2009}) might be used tune the electronic and vibrational structure of disordered crystals.\cite{Overy_2016} To some extent these relationships are already being used empirically to link \emph{e.g.}\ displacive and/or compositional disorder to phonon anomalies in thermoelectrics and ferroelectrics,\cite{Hlinka_2003,Delaire_2011} or to the spin-liquid behaviour of frustrated magnets.\cite{Zhang_2018} Yet there is much still to be understood regarding these disorder--property relationships in more general terms. A common theme is the intermediacy of collective behaviour; for example, electronic and vibrational states may be neither entirely localised nor extensive. In the domain of photonic materials, such intermediate states are responsible for a variety of fascinating optical properties,\cite{Wiersma_2013} and the long-term goal in this respect has to be exploitation of gauge field symmetries to engineer materials with equally anomalous electronic properties. Moreover, the gauge field excitations are themselves of great interest; while these have been studied extensively in the context of the magnetic monopoles of spin-ices,\cite{Castelnovo_2008} the analogous topological defects, their interactions, and their emergent dynamics in non-magnetic disordered crystals remain largely unexplored.

\noindent{\bf Unconventional data storage.} It was Schr{\"o}dinger who suggested that aperiodic crystals might be used to store information,\cite{Schrodinger_1944} and---in this spirit---Mackay highlighted that DNA can be considered a disordered 1D crystal with each unit (base) in one of four possible states.\cite{Cartwright_2012} Most conventional approaches to data storage follow variants on this same idea: each of $N$ bits is arranged on some periodic array (in 1, 2, or 3D) and encodes information by adopting one of some fixed number $n$ of degenerate states. Hence the total number of possible configurations is $n^N$. The information stored in each bit is independent of that in the others such that, if its contents are erased, the information is lost forever. Correlated disordered states offer an unconventional alternative that might prove useful in ensuring information redundancy. For disordered systems with extensive configurational entropies, the configurational landscape also grows as $n^N$, but $n$ is usually non-integral. This reflects a classical entanglement of local degrees of freedom across arbitrarily large distances within the crystal (related to the intermediate nature of collective states discussed above). A remarkable consequence is that the information contained within a single is now recoverable on erasure because of this inbuilt degeneracy [Figure 7]. Hence disordered crystals may provide a unique opportunity for high-density data storage and manipulation with intrinsic error-correcting capabilities.

\begin{figure}
\begin{center}
\includegraphics[width=8.3cm]{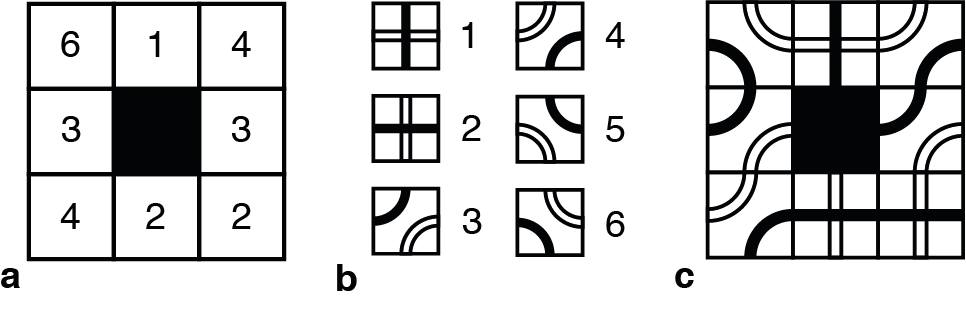}
\end{center}
\caption{\label{fig7} {\bf Data storage in disordered states.} (a) Information stored in a conventional array of bits (here 1--6) is irrecoverable if a single bit is erased (black square). (b) Mapping tiles to bit states; note each tile has two black edges and two white edges, \emph{cf}.\ the square-ice model of Ref.~\citenum{Lieb_1967}. That this model has an extensive entropy ensures that the number of possible tilings (satisfying matching rules) grows with system size, and hence such tilings offer a means of information storage. (c) The information encoded within each bit is no longer independent of its neighbours, such that the loss of one tile does not imply loss of information. Here, it is clear that the missing tile must correspond to state `5'.}
\end{figure}

{\bf Central challenges.} Whatever direction the field takes, we anticipate that four central challenges will play a crucially important role in developing new families of functional disordered crystals. The first is the task of establishing the underlying theory: what states are possible; how do they depend on degrees of freedom, interactions, and lattice geometries; what are their entropies, symmetries, excitations? Much is already known in this respect, but even ostensibly well-understood systems continue to surprise. The second challenge is the more chemical: how do we, as materials chemists, controllably introduce or manipulate a specific type of disorder within a crystalline material? Indeed the focus of this review has been on key strategies for achieving precisely this type of control, but of course our limited exposition cannot be the whole story. Third, there is the difficult experimental challenge of characterising the disorder present within any such crystal. The historical emphasis has been on entropy and spectroscopic measurements; more recently this emphasis has shifted to the application of scattering techniques, including pair distribution function and single-crystal diffuse scattering studies. In all cases, the difficulty lies not only in measurement but in interpretation. And, finally, there is the challenge of establishing a robust link between physical properties and the type of disorder present. Sometimes this link will be obvious (plastic crystals are soft, \emph{e.g.}), but we expect such cases are in the minority. Instead, solving this problem will more frequently rely on the application of computational methods, which for disordered systems is its own particular challenge.

\section*{Acknowledgements}

A.S. and A.L.G. gratefully acknowledge financial support from the E.R.C. (Grant 788144) and the Leverhulme Trust U.K. (Grant No. RPG-2015-292). A.S. thanks the Swiss National Science Foundation for Ambizione and PostDoc Mobility Fellowships (PZ00P2\_180035, P2EZP2\_155608). A.L.G. acknowledges many useful discussions with collaborators past and present.

\bibliography{nrc_2020_disorder} 

\end{document}